# Enhancement of ferromagnetism by nickel doping in the "112" cobaltite $EuBaCo_2O_{5.50}$

B. Raveau, Ch. Simon, V. Caignaert, V. Pralong and F.X. Lefevre

Laboratoire CRISMAT, ENSICAEN/CNRS, UMR6508, 6 Boulevard du Maréchal Juin, 14050 CAEN cedex 4, France
Tel: 33 (0)2 31 45 26 04, Fax: 33 (0)2 31 95 16 00
e-mail : bernard.raveau@ensicaen.fr

**Abstract**

The study of the ordered oxygen deficient perovskite $EuBaCo_{2-x}Ni_xO_{5.50}$ shows that the doping of cobalt sites by nickel induces a strong ferromagnetic component at low temperature in the antiferromagnetic matrix of $EuBaCo_2O_{5.50}$. This system exhibits indeed phase separation, i.e. consists of ferromagnetic domains embedded in the antiferromagnetic matrix of $EuBaCo_2O_{5.50}$. Besides, a magnetic transition is observed for the first time at 40K in the undoped and nickel doped phases, which can be attributed to the ferromagnetic ordering of the $Eu^{3+}$ moments below this temperature. Moreover sharp ultra magnetization multisteps are observed below 5K, characteristic of motion of domain walls in a "strong pinning" system and very different from any metamagnetic transition.

**Introduction**

Among the oxides which exhibit strong electron correlations, the cobaltites represent, beside the high Tc superconducting cuprates and the colossal magnetoresistance (CMR) manganites, the third class that is now investigated for its fascinating physics, made difficult to understand, due to the complexity of the spin, charge and orbital states that appear in these materials. After the discovery of the "112" ordered oxygen deficient perovskites $LnBaCo_2O_{5.4}$ (Ln = Eu, Gd), with original metal-insulator and ferro-antiferromagnetic transitions coupled with giant magnetoresistance [1], numerous studies were carried out to understand the attractive physics of the $LnBaCo_2O_{5+x}$ oxides structure [2-12]. In this series, the oxides corresponding to Ln = Eu, Gd and Tb exhibit a very similar behavior, due to the fact that the oxygen content corresponding to the ideal formula $LnBaCo_2O_{5.5}$, can be more easily controlled, just by synthesizing the compounds in air. Recently, Taskin et al [13] studied single crystals of $GdBaCo_2O_{5+x}$ and established the phase diagram of this system showing that its magnetic and transport properties are dominated by a nanoscopic phase separation either into two insulating phases in the electron doped region (for x < 0.50) or into one insulating and one metallic phase in the hole doped regions (for x > 0.50). Thus, it appears that the compounds $LnBaCo_2O_{5.5}$, which are sitting at the boundary between these two regions and contain only $Co^{3+}$ should exhibit a rather unstable distribution of the spins and carriers. The recent study of the spin structure of $TbBaCo_2O_{5.5}$, performed by Plakhty et al [14] shows indeed that the spin states of cobalt change in a very complex way versus temperature, so that the models of two-leg ladders of low spin and intermediate spin $Co^{3+}$ in pyramidal coordination separated by non magnetic $Co^{3+}$ octahedral layers cannot be used to interpret the magnetic and transport properties of $LnBaCo_2O_{5.50}$. Instead, an antiferromagnetic structure is clearly established for T ≤ 100K, where one of the $Co^{3+}$ species in octahedral coordination is in the high spin state, the rest being intermediate spin [14].

Bearing in mind the above results, it appears that the magnetic properties of the cobaltites $LnBaCo_2O_{5.5}$ should be very sensitive to the doping of the cobalt sites by various cations. We report herein on the magnetic properties of the cobaltites $EuBaCo_{2-x}Ni_xO_{5.5+\delta}$. We show that the substitution of nickel for cobalt induces a strong ferromagnetic component at low temperature (T ≤ 10K), and that simultaneously a new magnetic transition appears at 40K, due to the ferromagnetic ordering of the $Eu^{3+}$ species. Thus, this system is phase separated at 10K, and consists of ferromagnetic (FM) domains, embedded in an antiferromagnetic (AFM) matrix. Moreover, ultrasharp magnetization multisteps are observed at T < 10K by applying a magnetic field. The latter are due to the displacement of the domain walls inside the ferromagnetic domains. The pinning of such walls by nickel is discussed.

**Experiment**

The samples $EuBaCo_{2-x}Ni_xO_{5.5+\delta}$ were prepared from the mixtures of oxides $Eu_2O_3$, $Co_3O_4$ and NiO and carbonate $BaCO_3$ in the stoichiometric proportions. After preliminary heating at 1000°C in air for 12 hrs, the mixtures were ground and then pressed in the form of parallelepiped bars and sintered at 1100°C for 12 hrs in air. They were finally cooled down to room temperature in 6 hrs.

The quality of the crystallization was checked by X ray diffraction, confirming the usual pattern, involving an orthorhombic cell $a_p \times 2a_p \times 2a_p$, characteristic of the "112" type structure previously observed for the oxides $LnBaCo_2O_{5.5}$ [4]. The energy dispersive spectroscopy analysis (EDS) performed with a kevex analyzer mounted on a JEOL 200CX electron microscope allowed the cationic composition "$EuBaCo_{2-x}Ni_x$" to be confirmed for x = 0.08, 0.10 and x = 0.15. Oxygen content was determined by iodometric titration as 5.47, 5.55 and 5.57 for x = 0.08, 0.10 and 0.15 respectively, leading to cobalt valences ranging from 3.01 to 3.15.

For comparison, a sample of "undoped" cobaltite $EuBaCo_2O_{5.5\pm\delta}$ was prepared. Its synthesis in air was carried out in the same conditions as described above for the Ni-doped phase. The so-obtained phase was found to be practically stoichiometric "$0_{5.50}$", corresponding to a cobalt valence close to +3, in the limits of the accuracy of the titration method.

The dc magnetization was measured by an extraction method in a Quantum Design PPM System up to 9T. Different speed variations of the ramping magnetic field were used in this experiment. The reproducibility of the results and particularly the magnetization steps reproducibility were checked in different samples prepared in the same manner. The data presented here are only a small amount of the measured data. However, the sample preparation method is very important to get reproducible results.

**Results and discussion**

The magnetization of zero filed cooled samples registered versus temperature shows complex magnetic transitions between 300K and 150K, similar to those previously observed for $GdBaCo_2O_{5.5}$, i.e. corresponding to paramagnetic (PM) – ferromagnetic (FM)-antiferromagnetic (AFM) transitions [1-9], which were recently studied in details on single crystals of $GdBaCo_2O_{5.5}$ by Taskin et al [12], in terms of competition between ferromagnetism and antiferromagnetism. On figure1, we have reported the magnetization of $EuBaCo_{2-x}Ni_xO_{5.5+\delta}$ for x=0 and x=0.15 at $10^{-2}$ T and 1.45T. At $10^{-2}$ T, the susceptibility presents, for both compounds a double transition at 40K and 240K respectively. The transition at 40K is very similar in both, the undoped and Ni-doped compounds. The transition at 240K presents a much smaller peak in the Ni-doped sample. The curves obtained at 1.45 T are quite surprising and it is very important for their interpretation to examine the hysterisis loops presented at 10K in figure 2. The x=0 compound exhibits a very small loop mainly representative of an antiferromagnetic compound with a small ferromagnetic component. The strange shape of the loop arises from a possible paramagnetic component already reported in a previous study [15]. The origin of this component, which remains when x is increased, is still unknown.

It is quite clear on figure 2 that the ferromagnetic component is increased very strongly as x increases, to reach $0.4\mu_B$/f.u. in x=0.15. The pinning of the domains walls remains quite strong since a magnetic field of 3T is necessary to quit the antiferromagnetic branch obtained in ZFC. In other words, the ferromagnetic component is completely hidden by the hysteresis at 10K up to 3T. This explains the shape of figure 1b: at low temperature, only the antiferromagnetic component can be observed even at 1.45T. In order to measure the ferromagnetic fraction, it is necessary to increase the magnetic field up to 5T and then to decrease it back to low field. This is presented on figure 3 which shows that the Ni doping induces a strong ferromagnetic component at low temperature, but decreases the ferromagnetic component around 240K. This is summarized in figure 4: On figure 4(a), the antiferromagnetic component is presented at 10K, extracted from the ZFC magnetization at 1.45T. The small part attributed to $Eu^{3+}$ is removed and the antiferromagnetic fraction is supposed to be proportional to the measured moment with an unknown proportionality factor. On figure 4(b), the ferromagnetic component is extracted from the remnant magnetization at 10K (after 5T). This shows that Ni doping increases the ferromagnetic component and decreases the antiferromagnetic one.

The origin of the ferromagnetic component observed at low temperature (below 50K) in the x=0 compound has probably to be found in the ferromagnetic ordering of the $Eu^{3+}$ moments. Indeed, in $GdBaCo_2O_{5.5}$, the $Gd^{3+}$ moments are not ordered [12] and a strong paramagnetic component is observed. In addition, the $Eu^{3+}$ ion is known to be with such a small moment (0.08$\mu_B$/f.u.) [16]. If one subtracts this ferromagnetic $Eu^{3+}$ component the additional ferromagnetic component is roughly proportional to x (figure 4).

The second interesting feature concerns the nature of ferromagnetism that appears with Ni doping. In order to choose between two possible classical interpretations (canting versus "phase separation"), we have performed low temperature (2K) hysteresis loops. Figure 5 shows the magnetic field dependence of the magnetization of $EuBaCo_{1.85}Ni_{0.15}O_{5.57}$ registered at 2 K, with a sweep rate of $5.10^{-3}$ T/sec. One observes very sharp magnetization steps as H increases and decreases and we notice that the demagnetization steps at the 1st descent are approximately symmetric with respect to the magnetization steps at the 2nd rising of the magnetic field, i.e. appear at close values of $|H|$, i.e., at 1T(0.1T), 3.3T(-3.4T), 5.1T(-4.9T), 7.2T(-7.3T). These results suggest that there is a possible phase separation at low temperature, i.e. that ferromagnetic (FM) domains are growing inside the antiferromagnetic (AFM) matrix. The appearance of such sharp magnetization multisteps may have different origins:

(i) spreading of the ferromagnetic domains at the expense of the antiferromagnetic matrix.
(ii) increase of the ferromagnetic moment by wall motion inside the ferromagnetic domains

The first hypothesis is based on the observations previously made on manganites doped with various elements [17-19]. Like our cobaltites, the manganites also exhibit phase separation involving ferromagnetic domains embedded in an antiferromagnetic matrix. It was shown that the multistep like magnetization behavior of these manganites was mainly governed by a martensitic-like mechanism involving a concerted motion of the atoms at the boundary between the FM and AFM phases, which do not exhibit the same crystallographic symmetry of their perovskite structure. In fact, the behavior of the present cobaltites is significantly different from that of manganites: in the cobaltites the magnetization obtained at decreasing field does not directly go back to zero, when the field is switched off, contrary to the manganites, but shows a significant remnant value and differently, sharp demagnetization steps are observed. Moreover, the coercive field that is obtained for such cobaltites is rather high. Thus, a second scenario, should be considered, which is more likely and deals with the wall motion inside the FM domains which should be large enough for that (micrometric size). Multistep magnetization reversal has indeed been previously observed in intermetallic ferromagnets [20-22] and interpreted by the different authors as domain wall motion, mainly based on pinning effects introduced by foreign atoms. The latter induce fluctuations in the exchange interaction and local crystal fields [20] which are at the origin of the pinning of the narrow domain walls. More recently Mushnikov et al [23], showed that such multisteps, observed in $Dy(Fe, M)_2$ with M = Al, Si, were due to sharp increase of the temperature under heat released at domain wall motion. In the cobaltite $EuBaCo_{1.85}Ni_{0.15}O_{5.57}$, there is no doubt that $Ni^{2+}$ species play a primordial role in the appearance of such multisteps at 2K. The hysteresis loop registered at 2K with a sweep rate of $5.10^{-3}$T/sec. for the virgin sample $EuBaCo_2O_{5.50}$ (Figure 6) clearly shows that the steps have practically disappeared. Thus, the $Ni^{2+}$ species change dramatically the local crystal field and induce a very effective pinning of the narrow domain walls at low temperature, responsible for the very sharp and large steps observed for the oxide $EuBaCo_{1.85}Ni_{0.15}O_{5.57}$ (figure 5). In contrast, in $EuBaCo_2O_{5.50}$, only very small steps are observed (figure 6) which may be due to the pinning by $Co^{4+}$ species. It is quite remarkable that the number of steps and the value of the critical field corresponding to the magnetization steps decrease as T increases, exactly as for the intermetallic phases $Dy(Fe, M)_2$ [23]. For instance, for $EuBaCo_{1.85}Ni_{0.15}O_{5.57}$ at T = 5K, and for a sweep rate of $5.10^{-3}$T/s., one observes only one large sharp step at +4.1T, at the second rising of the field, whereas very small steps remain around + 0.9T - 1T (figure 7). Finally at 10 K all the steps have practically disappeared (figure 2). It is also worth pointing out that the position of the steps (the value of the critical field corresponding to the steps) and the height of the steps depend on the sweep rate. Figure 8 shows the hysteresis loops registered at 2K for $EuBaCo_{1.85}Ni_{0.15}O_{5.57}$ for three different sweep rates $5.10^{-3}$T/s., $10^{-2}$T/s. and $2.5.10^{-2}$T/s. One observes that the amplitude of the steps decreases significantly as the sweep rate decreases, in agreement with the results observed for the intermetallic phases $DyFe_2$ doped with Al or Si [23]. In contrast, the value of the critical field corresponding to the steps, increases with the sweep rate, contrary to what has been observed for $DyFe_2$ [23], as well as for manganites [17-19].

The origin of the ferromagnetic component in this "112" series cannot be explained in a simple way. In the undoped phase (x=0), the oxygen stoichiometry of this oxide $EuBaCo_2O_{5.50}$, which imposes the only presence of trivalent cobalt, suggests that only $Co^{3+}$-O-$Co^{3+}$ interactions would be possible. Thus, the small ferromagnetic component below 40K observed for this oxide cannot be explained by such interactions, which should be antiferromagnetic according to Kamamori-Goodenough rules [24], and rather originates from the ordering of europium moments. Nevertheless, this oxide (figure 9a), built up $CoO_5$ pyramids and $CoO_6$ octahedra may favor a small disproportionation of trivalent cobalt, according to the equation $2Co^{3+} \to Co^{4+}+Co^{2+}$, so that $Co^{2+}$ would preferentially sit in the $CoO_5$ pyramids and $Co^{4+}$ in the $CoO_6$ octahedra. As a consequence, ferromagnetic $Co^{3+}$-O-$Co^{4+}$ interactions are also quite possible in this oxide, even if the global valence of cobalt is +3. In the Ni-doped phases, the significant enhancement of the ferromagnetic domains by nickel doping is also not compatible, at first sight with Kamamori-Goodenough rules for a simple perovskite system, since the $Ni^{2+}$-O-$Co^{3+}$ interactions should be strongly antiferromagnetic. In fact, this particular ordered oxygen deficient perovskite structure (figure 9a) may explain easily these FM interactions. If one admits that the AFM structure of the ideal stoichiometric matrix $EuBaCo_2O_{5.50}$ at 10K is similar to that observed for $TbBaCo_2O_{5.50}$ as determinated by Plakhty et al [14], the latter can be described (figure 9b) in the following way. A perfect "112" AFM matrix consists of planes built up of double antiferromagnetic rows which alternate with double ferromagnetic rows. In each row one $Co^{3+}$ cation in pyramidal coordination alternates with one $Co^{3+}$ cation in octahedral coordination. The two adjacent ferromagnetic rows are themselves antiferromagnetically coupled. The entire magnetic structure is then made by the stacking of such identical $Co^{3+}$ planes, two successive planes being antiferromagnetically coupled. When one $Ni^{2+}$ species is introduced in the structure, it will sit on an octahedral site and will couple antiferromagnetically with its next cobalt neighbors (figure 9c) and will create in this way $[Ni^{2+}(Co^{3+})_6]$ ferrimagnetic octahedral clusters. Moreover, we observe that simultaneously to nickel substitution, $Co^{4+}$ content is increasing, so that $Co^{4+}$-O-$Co^{3+}$ and $Co^{4+}$-O-$Ni^{2+}$ ferromagnetic interactions may develop, which can also be closely related to the ferrimagnetic clusters.

**Conclusion**

This study of the substitution of nickel for cobalt in the "112" type cobaltite $EuBaCo_2O_{5.50}$ shows that the introduction of $Ni^{2+}$ on the cobaltite sites induces the formation of a strong FM component at low temperature inside the initial AFM matrix of the ideal stoichiometric cobaltite $EuBaCo_2O_{5.50}$. Thus, this system is phase separated below 40K, i.e. consists of FM domains embedded in the AFM matrix of $EuBaCo_2O_{5.50}$. At low magnetic field in ZFC mode, the material appears antiferromagnetic, due to the fact that the magnetic moments inside the FM domains are compensated. Due to the particularly complex structure of the AFM phase, the FM domains cannot be extended to the whole sample. The existence of a small FM component below 40K in the case of the pure $EuBaCo_2O_{5.50}$ phase can be explained by the ferromagnetic ordering of $Eu^{3+}$. In contrast, the enhancement of the FM fraction by Ni doping cannot be explained by $Ni^{2+}$-O-$Co^{3+}$ interactions which should be strongly antiferromagnetic. More likely it results from the formation of octahedral ferrimagnetic clusters around each nickel, combined with $Ni^{2+}$-O-$Co^{4+}$ and $Co^{3+}$-O-$Co^{4+}$ ferromagnetic interactions.

The second important point of this study concerns the evidence for ultrasharp magnetization multisteps, that are induced by nickel doping at very low temperature. It is the first time that such multisteps are observed in the cobaltites. In spite of the existence of phase separation in these oxides, it appears clearly that the phenomenon that is observed for cobaltites is very different from the martensitic mechanism previously shown for the manganites. Instead, a scenario similar to that observed in intermetallic ferromagnets, involving a very effective pinning of the narrow domain walls inside the FM domains by $Ni^{2+}$ species can be proposed. The latter originates most likely from the local structural distortions induced by $Ni^{2+}$, and from its combination with the presence of $Co^{4+}$, leading to fluctuations in the local crystal field and consequently in exchange interactions. Further investigations, controlling the oxygen stoichiometry, but also the nature of impurities will be necessary to understand these phenomena.

**Figure captions**

Figure 1: Temperature dependence of the magnetization of $EuBaCo_2O_{5.50}$ and $EuBaCo_{1.85}Ni_{0.15}O_{5.57}$ registered in (a) $10^{-2}$ T and (b) 1.45T. The samples were zero field cooled before measurements.

Figure 2: Hysteresis loops registered at 10 K for (a) $EuBaCo_2O_{5.50}$, (b) $EuBaCo_{1.92}Ni_{0.08}O_{5.47}$, (c) $EuBaCo_{1.90}Ni_{0.1}O_{5.5}$, (d) $EuBaCo_{1.85}Ni_{0.15}O_{5.57}$.

Figure 3: Temperature dependence of the magnetization of $EuBaCo_2O_{5.50}$ and $EuBaCo_{1.85}Ni_{0.15}O_{5.57}$ after increase / decrease of the magnetic field. The samples were first zero field cooled (5K), then the field was increased up to 5 T and finally decreased to $10^{-2}$ T and M(T) was registered at this latter value of the field.

Figure 4: Magnetic components of the oxides $EuBaCo_{2-x}Ni_xO_{5.5+\delta}$ at 10 K extracted from the zfc magnetization in 1,45T (a) and from the remnant magnetization after 5 T (b).

Figure 5: Magnetic field dependence of the magnetization of $EuBaCo_{1.85}Ni_{0.15}O_{5.57}$ registered at 2 K, with a sweep rate of $5.10^{-3}$ T/s.

Figure 6: Magnetic field dependence of the magnetization of $EuBaCo_2O_{5.50}$ registered at 2 K, with a sweep rate of $5.10^{-3}$ T/s.

Figure 7: Magnetic field dependence of the magnetization of $EuBaCo_{1.85}Ni_{0.15}O_{5.57}$ registered at 5 K, with a sweep rate of $5.10^{-3}$ T/s.

Figure 8: Magnetic field dependence of the magnetization of $EuBaCo_{1.85}Ni_{0.15}O_{5.57}$ registered at 2 K for three sweep rates of (a) $5.10^{-3}$ T/s (b) $10^{-2}$ T/s (c) $2.5.10^{-2}$ T/s.

Figure 9: Schematized nuclear (a) and magnetic (b) structure of $EuBaCo_2O_{5.50}$ at 10 K and doping by $Ni^{2+}$ (c) forming octahedral $[Ni^{2+}Co_6^{3+}]$ clusters.

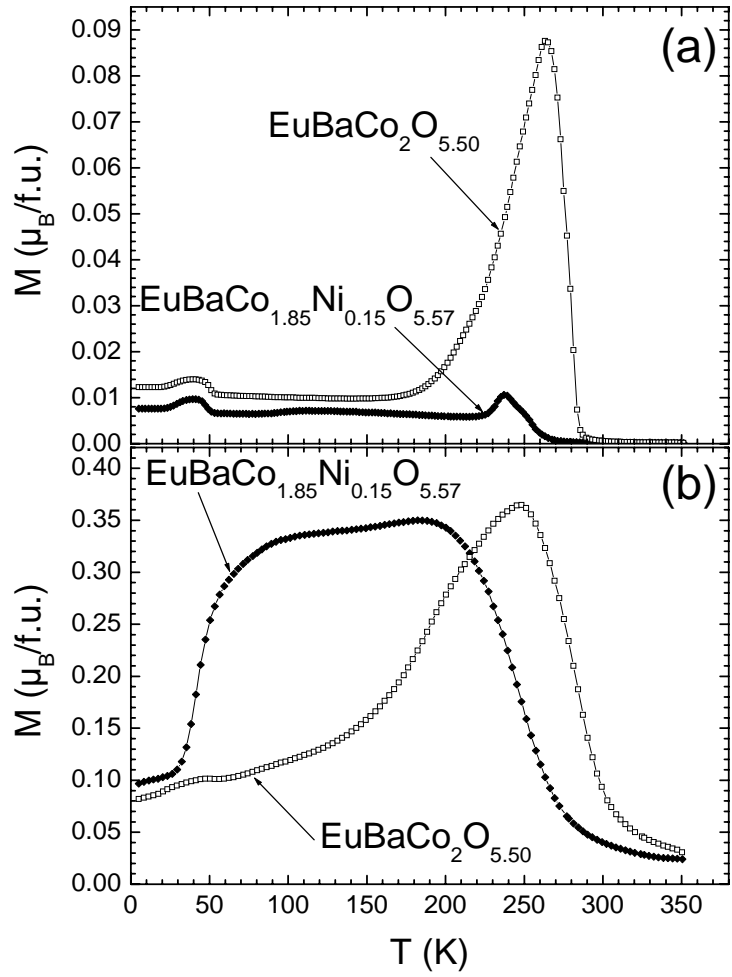

Temperature dependence of the magnetization of EuBaCo$_2$O$_{5.50}$ and EuBaCo$_{1.85}$Ni$_{0.15}$O$_{5.57}$ registered in (a) 10$^{-2}$ T and (b) 1.45T. The samples were zero field cooled before measurements.

**Fig. 1**

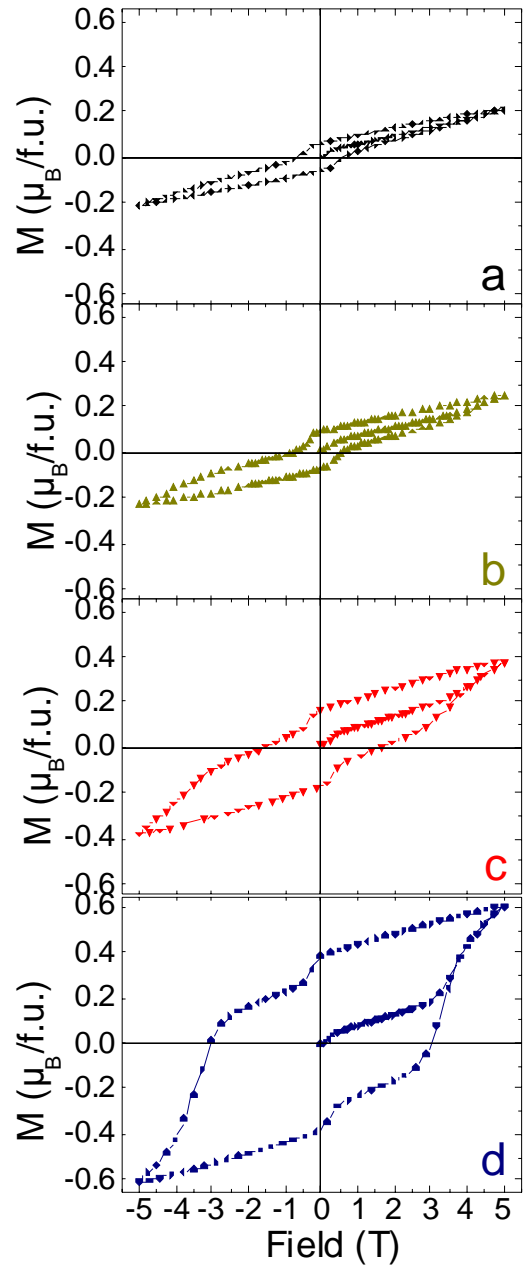

Hysteresis loops registered at 10 K for (a) $EuBaCo_2O_{5.50}$, (b) $EuBaCo_{1.92}Ni_{0.08}O_{5.47}$, (c) $EuBaCo_{1.90}Ni_{0.1}O_{5.5}$, (d) $EuBaCo_{1.85}Ni_{0.15}O_{5.57}$.

**Fig. 2**

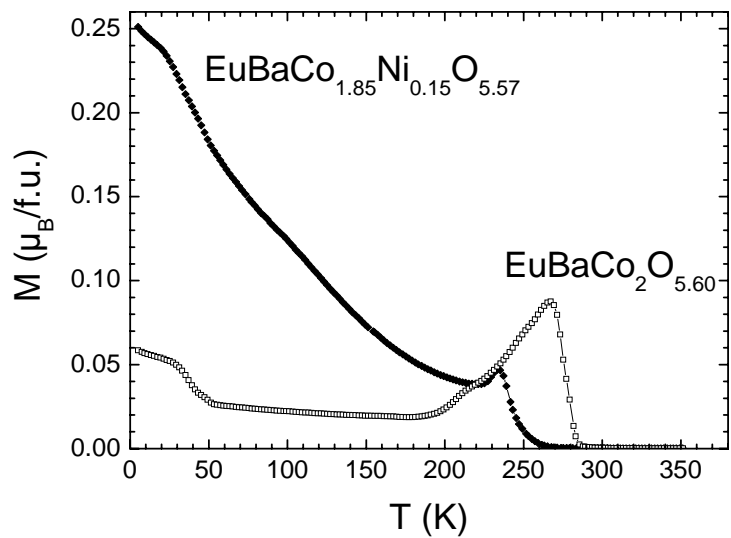

Temperature dependence of the magnetization of $EuBaCo_2O_{5.50}$ and $EuBaCo_{1.85}Ni_{0.15}O_{5.57}$ after increase / decrease of the magnetic field. The samples were first zero field cooled (5K), then the field was increased up to 5 T and finally decreased to $10^{-2}$ T and M(T) was registered at this latter value of the field.

**Fig. 3**

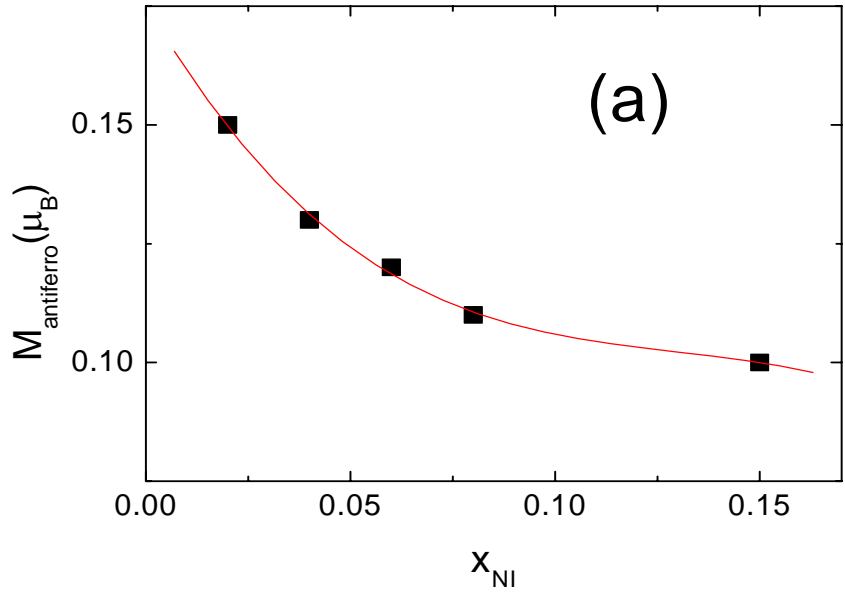

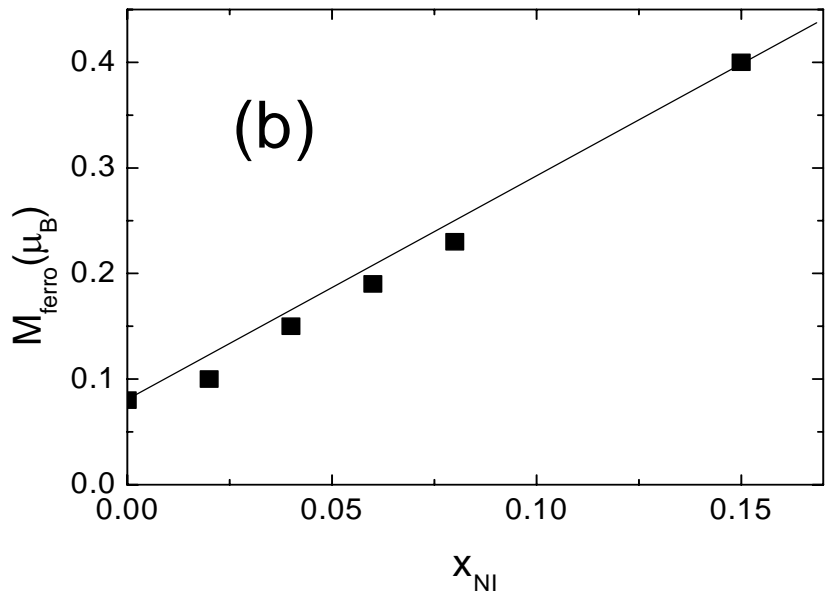

Magnetic components of the oxides EuBaCo$_{2-x}$Ni$_x$O$_{5.5+\delta}$ at 10 K extracted from the zfc magnetization in 1,45T (a) and from the remanent magnetization after 5 T (b).

Fig. 4

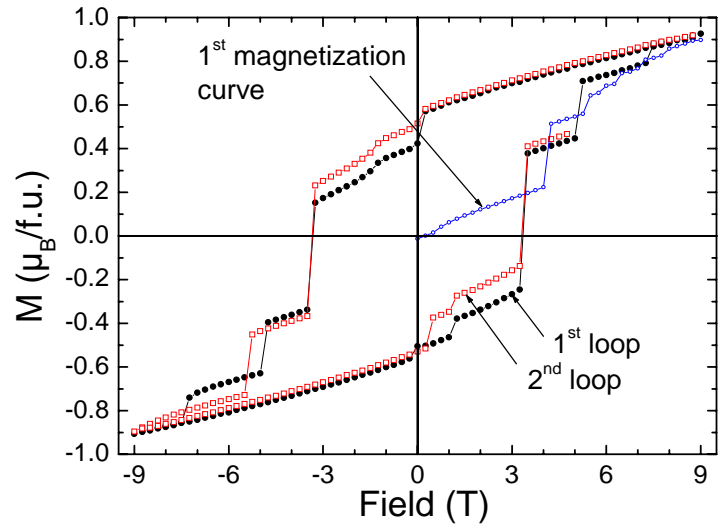

Magnetic field dependence of the magnetization of EuBaCo$_{1.85}$Ni$_{0.15}$O$_{5.57}$ registered at 2 K, with a sweep rate of $5.10^{-3}$ T/s.

**Fig. 5**

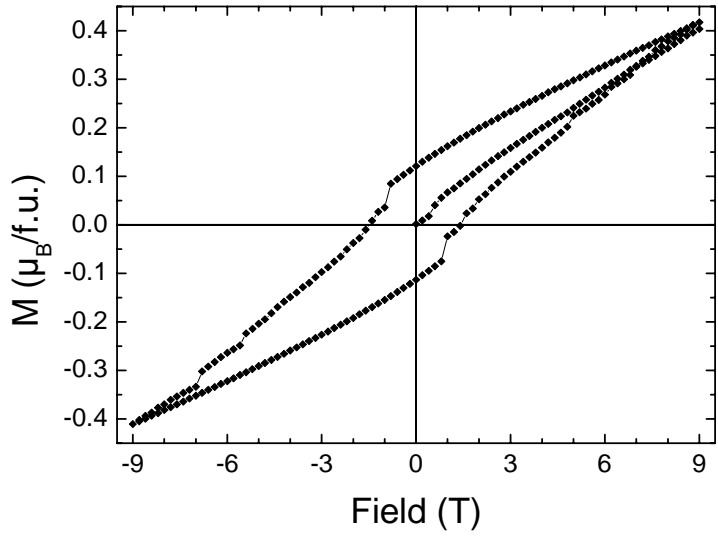

Magnetic field dependence of the magnetization of EuBaCo$_2$O$_{5.50}$ registered at 2 K, with a sweep rate of $5.10^{-3}$ T/s.

**Fig. 6**

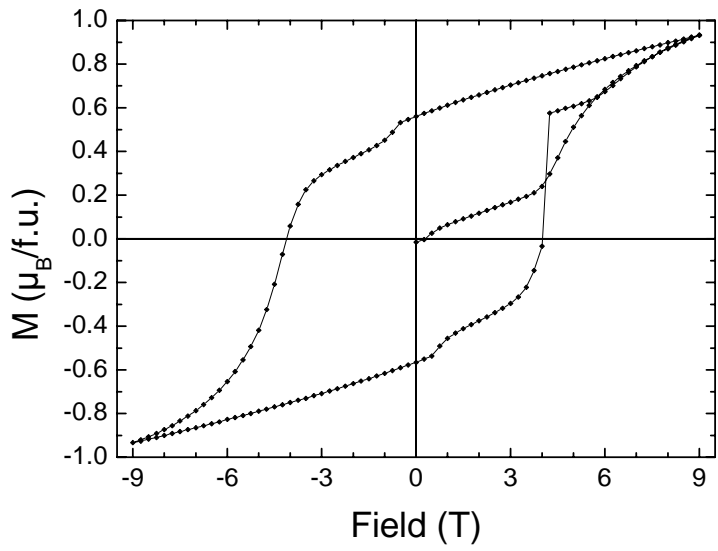

Magnetic field dependence of the magnetization of EuBaCo$_{1.85}$Ni$_{0.15}$O$_{5.57}$ registered at 5 K, with a sweep rate of 5.10$^{-3}$ T/s.

**Fig. 7**

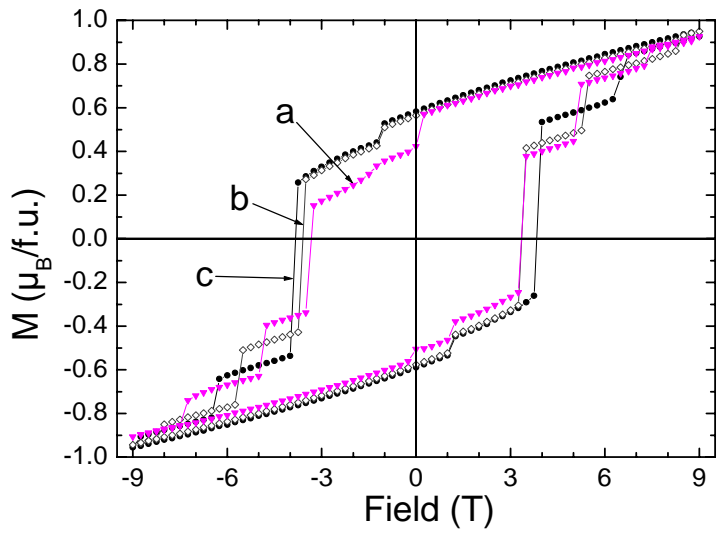

Magnetic field dependence of the magnetization of EuBaCo$_{1.85}$Ni$_{0.15}$O$_{5.57}$ registered at 2 K for three sweep rates of (a) $5.10^{-3}$ T/s (b) $10^{-2}$ T/s (c) $2.5.10^{-2}$ T/s.

**Fig. 8**

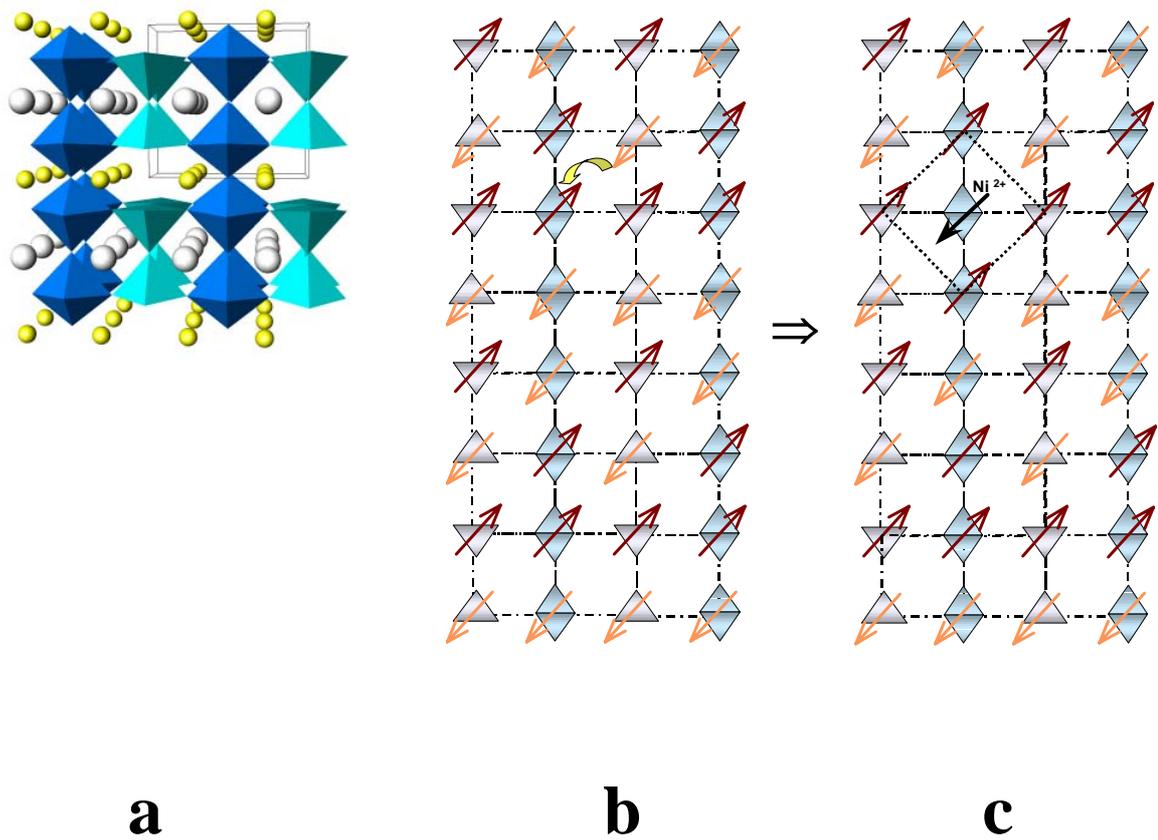

**a**           **b**           **c**

Schematized nuclear (a) and magnetic (b) structure of $EuBaCo_2O_{5.50}$ at 10 K and doping by $Ni^{2+}$ (c) forming octahedral $[Ni^{2+}Co_6^{3+}]$ clusters.

**Fig. 9**